\documentclass[prb,twocolumn,showpacs,preprintnumbers,amsmath,amssymb]{revtex4}

\usepackage{dcolumn}
\usepackage{bm}
\usepackage{graphicx}

\begin{document}


\title{Influence of magnetic field offsets on the resistance of magnetic barriers in two-dimensional electron gases}

\author{S. Hugger}\author{M. Cerchez}\author{H. Xu}\author{T. Heinzel}
\email{thomas.heinzel@uni-duesseldorf.de}
\affiliation{Heinrich-Heine-Universit\"at, Universit\"atsstr. 1,
40225 D\"usseldorf, Germany}
\date{\today}

\begin{abstract}
Magnetic barriers in two-dimensional electron gases are shifted in B space by homogeneous, perpendicular magnetic fields. The magnetoresistance across the barrier shows a characteristic asymmetric dip in the regime where the polarity of the homogeneous magnetic field is opposite to that one of the magnetic barrier. The measurements are in quantitative agreement with semiclassical simulations, which reveal that the magnetoresistance originates from the interplay of snake orbits with $E\times B$ drift at the edges of the Hall bar and with elastic scattering.
\end{abstract}

\maketitle

\section{\label{sec:1}INTRODUCTION}

The transport properties of two-dimensional electron gases (2DEGs) in spatially varying magnetic fields show a rich phenomenology, and several interesting effects have been reported recently. For example, magnetic superlattices
\cite{Ye1995,Nogaret1997} show commensurability oscillations in the magnetotransport, while giant magnetoresistance effects have been found on one-dimensional magnetic arrays \cite{Nogaret1997}. Also, a variety of single magnetic nanostructures have been investigated via their influence on the 2DEG, like resistance resonances along magnetic edge states \cite{Nogaret2000}, the experimental realization of magnetic waveguides \cite{Nogaret2003}, or the demonstration of Hall sensing \cite{Peeters1998,Novoselov2002}. A magnetic nanostructure of elementary character, known as \emph{single magnetic barrier - MB}, is formed by the highly localized perpendicular ($z$-) component $B_{fz}$ of the magnetic fringe field in a 2DEG below the edge of a ferromagnetic film magnetized in transport ($x$-) direction  \cite{Peeters1993,Matulis1994,Monzon1997,Johnson1997,Kubrak2000,Vancura2000,Gallagher2001,Kubrak2001,Cerchez2007}. Alternatively, such a structure can be defined in a 2DEG with a graded step \cite{Leadbeater1995}.
Various aspects of MBs have been studied in the past few years. The height of the MB can be conveniently tuned by in-plane magnetic fields, while their magnetoresistance is interpreted in a semiclassical picture \cite{Monzon1997,Johnson1997,Kubrak2000,Vancura2000,Gallagher2001,Kubrak2001,Hong2002,Cerchez2007}, within which electrons experience a deflection by $B_{fz}$, such that for a fixed Fermi energy, the angle of incidence determines whether the electrons get transmitted or reflected \cite{Peeters1993,Ibrahim1997}. MBs in quantum wires, on the other hand, have attracted a lot of recent attention from theory. These studies have established the potential of MBs as tunable spin filters \cite{Majumdar1996,Guo2000,Papp2001a,Papp2001b,Xu2001,Lu2002,Guo2002,Jiang2002,Xu2005,Zhai2005,Zhai2006} and predict the presence of Fano resonances \cite{Xu2007}.\\
In all experiments reported so far, the resistance of MBs has been investigated as a function either of a magnetic field applied in x-direction, or of the carrier density \cite{Leadbeater1995,Monzon1997,Johnson1997,Kubrak2000,Vancura2000,Gallagher2001,Kubrak2001,Cerchez2007}. Therefore, the polarity of the total magnetic field in z - direction $B_{z}(x)$ is constant. A qualitatively new situation arises when $B_{z}(x)$ changes its sign across the MB, since novel types of trajectories with a snake orbit character become possible.\\
In the present paper, we report the experimental implementation of such a magnetic barrier structure. It is generated by superimposing a homogeneous perpendicular magnetic field $B_{hz}$ to $B_{fz}(x)$. This way, a MB of constant strength and shape is displaced along the $B_z$ - axis. The most interesting situation arises when $B_{hz}$ partly compensates $B_{fz}(x)$, such that lines of zero magnetic field along the y-direction exist, which act as guiding centers for snake orbits. In this range, a pronounced dip of asymmetric shape in the magnetoresistance is observed. We explain this structure by an interplay of the conductance enhancement along y direction due to snake orbits with both elastic scattering as well as $E\times B$ drifts at the edges of the Hall bar. Numerical simulations based on the Kubo formalism and the Landauer-B\"uttiker model substantiate this interpretation.\\
The outline of the paper is as follows: Section II is dedicated to the experimental results. An interpretation of the data with the help of simulations is given in Section III, followed by a summary in Section IV.

\section{\label{sec:2}EXPERIMENTS}

\begin{figure}
\includegraphics{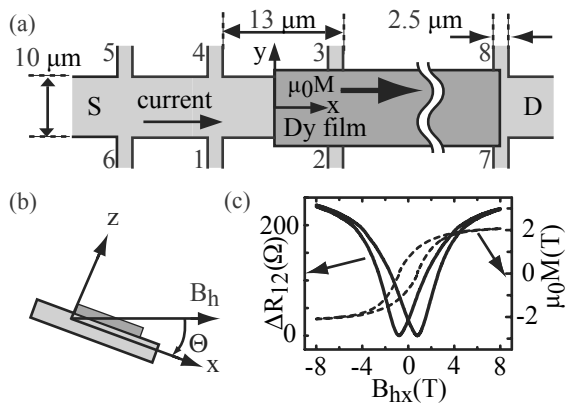}
\caption{(a) Top view of the schematic sample: a Dy platelet is placed on top of a Hall bar with source (S) and drain (D) contacts as well as voltage probes 1 to 8. (b) Nomenclature used for rotated samples. (c) The resistance $\Delta R_{12}(B_{hx})$ of the MB (the background resistance of the 2DEG is subtracted), measured at a temperature of $2\,\mathrm{K}$, and the corresponding magnetization of the Dy film.} \label{fig:1}
\end{figure}

\begin{figure}
\includegraphics{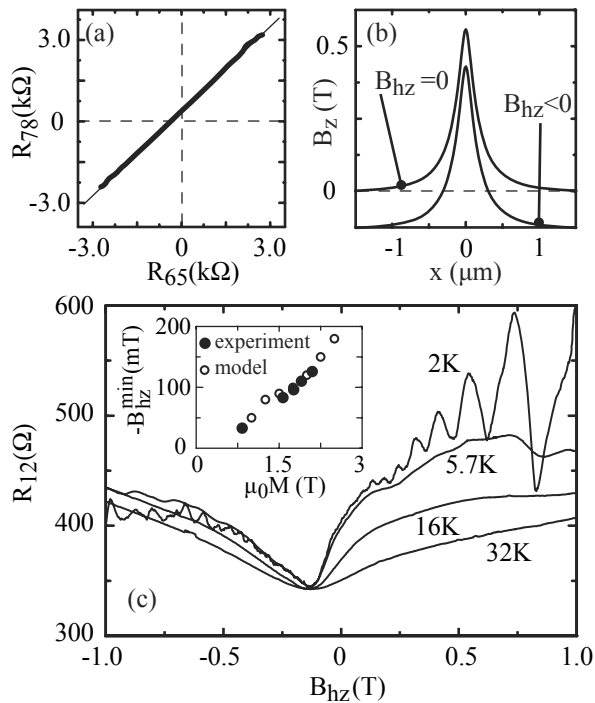}
\caption{(a) The Hall resistance $R_{78}$ vs. $R_{65}$, measured at $2\,\mathrm{K}$ (bold line) and the linear fit to the data (thin line), with a slope of 1.04. (b) Sketch of $B_z(x)=B_{fz}(x)+B_{hz}$ for $B_{hz}=0$ and $B_{hz}=-125\,\mathrm{mT}$.  (c) Measurements of $R_{12}(B_{hz})$. Inset: measured and simulated values of $B_{hz}^{min}$ as a function of the film magnetization.}
\label{fig:2}
\end{figure}

The experiments have been performed on a commercially available $\mathrm{GaAs/Al_xGa_{1-x}As}$ - heterostructure \cite{IntelliEpi} with a 2DEG $65\,\mathrm{nm}$ below the surface. The sample geometry is shown in Fig. \ref{fig:1}(a). The samples were prepared by optical lithography, followed by wet chemical etching for the definition of the Hall bar, or by metallization steps, respectively. Au/Ge Ohmic contacts were defined at source and drain contacts and at the voltage probes. A dysprosium (Dy) platelet with a thickness of $300\,\mathrm{nm}$ was deposited on the heterostructure surface by thermal evaporation at a base pressure of $2\times 10^{-6}\,\mathrm{mbar}$.  A Cr/Au layer of $150\,\mathrm{nm}$ thickness was deposited on top to prevent the Dy from oxidizing under ambient conditions. The samples were measured in a liquid helium cryostat with a variable temperature insert. A rotatable sample stage allows orientation of the sample within the x-z plane, see Fig. \ref{fig:1}(b), with an accuracy of the rotation angle $\Theta$ of $0.1$ degrees. The cryostat is equipped with a superconductive magnet with a maximum homogeneous magnetic field of $B_h=8\,\mathrm{T}$.
A current of $I=100\,\mathrm{nA}$ with a frequency of $13\,\mathrm{Hz}$ was passed from source to drain. Conventional four-probe measurements as a function of perpendicular magnetic field $B_{hz}$ were performed, from which an electron density of $2.45 \times 10^{15}\,\mathrm{m^{-2}}$ and a mobility of $22.1\,\mathrm{m^2/Vs}$ (corresponding to a Drude scattering time of $\tau_D = 8.4\,\mathrm{ps}$ and an elastic mean free path of $1.8\,\mathrm{\mu m}$) were obtained at a temperature of $2\,\mathrm{K}$. The quantum scattering time was determined from the envelope of the Shubnikov - de Haas oscillations \cite{Ando1982} as $\tau_q =1.04\,\mathrm{ps}$ to an accuracy of 10\%. The vanishing of the Hall voltage was used to adjust $\Theta =0$.\\
In a previous publication, we have discussed in detail the resistance of such a MB as a function of a parallel magnetic field $B_{hx}$ \cite{Cerchez2007}. The fringe field $B_{fz}(x)$ is given by \cite{Ibrahim1997,Vancura2000}
\begin{equation}
B_{fz}(x)=-\frac{\mu_0M(B_{hx})}{4\pi}\ln{\left(\frac{x^2+z_0^2}{x^2+(z_0+h_0)^2}\right)} \label{eq:1}
\end{equation}
see Fig. \ref{fig:2}(a). Here, $z_0$ is the distance of the 2DEG from the surface and $h_0$ denotes the thickness of the Dy film with a magnetization denoted by $M(B_{hx})$.\\
The MB was characterized by measuring its resistance $R_{12}$ as a function of $B_{hx}$ at $\Theta = 0$. Furthermore, from the fit of $R_{12}(B_{hx})$ to a semiclassical model \cite{Cerchez2007}, the magnetization characteristics of the Dy film at its edge of relevance is obtained. Five samples were measured, all with qualitatively identical behavior. Here, we focus on the sample with the strongest MB of $B_{fz}^{max}\equiv B_{fz}(x=0)= 0.57\,\mathrm{T}$ for $B_{hx}=8\,\mathrm{T}$, corresponding to a saturation magnetization of $\mu_0M_{s} =2.1\,\mathrm{T}$.

With the MB established this way, we superimpose a homogeneous, perpendicular magnetic field $B_{hz}$. For this purpose, the sample is rotated at $B_h=8\,\mathrm{T}$ by small angles, such that the Dy film is kept at constant magnetization, while a homogeneous magnetic field in z direction $B_{hz}$ emerges, see Fig. \ref{fig:1}(b). In Fig. \ref{fig:2}(a), $R_{78}(B_{hz})$ is plotted as a function of $R_{65}(B_{hz})$. A linear relation with a slope of $dR_{78}/dR_{65}= 1.04$ and an offset of $380\,\mathrm{\Omega}$ is found. Since both Hall crosses measure the average magnetic field in the probed section, we conclude that the MB is not modified by rotations in the range of interest, namely for $|B_{hz}|\leq 1 \,\mathrm{T}$, corresponding to $|\Theta|\leq 7.2^{\circ}$. Thus, rotating the sample in this range corresponds to a displacement of $B_{fz}(x)$ along the $B_z$ - axis, as sketched in Fig. \ref{fig:2}(b). Alternatively, we turned $B_h$ off and set $\Theta = 90^{\circ}$, in which case the MB originates from the remanence of the Dy film. In that setup, $|B_h|=|B_{hz}| \leq 0.2\,\mathrm{T}$ can be applied without changing the film magnetization, which is kept in-plane by the shape anisotropy.\\
In Fig. \ref{fig:2}(c), the magnetoresistance traces $R_{12}(B_{hz})$ at various temperatures are reproduced. A pronounced asymmetric dip of $\approx 90\,\mathrm{\Omega}$ strength at $2\,\mathrm{K}$, located at $B_{hz}^{min}=-125\,\mathrm{mT}$, is observed. As the temperature increases, the dip broadens, both its amplitude and its asymmetry are reduced, and the position of the minimum shifts slightly towards $B_{hz}=0$, but the structure remains clearly visible at $32\,\mathrm{K}$. The weak temperature dependence of the structure indicates a classical origin. In the inset of Fig. \ref{fig:2}(c), $B_{hz}^{min}$ is plotted as a function of the film magnetization, which is tuned by $B_{hx}$. $B_{hz}^{min}$ drops approximately linearly as the magnetization is reduced. Furthermore, we notice a suppression of the Shubnikov - de Haas oscillations for $B_{hz}<0$.

\section{\label{sec:3}SIMULATION AND INTERPRETATION}

An insightful, yet only qualitative, interpretation of the magnetoresistance is possible within the Kubo formalism \cite{Kubo1957}, which relates the components $\sigma_{ij}, (\{i,j\}=\{x,y\})$ of the conductivity tensor to the electron velocity correlation functions $\langle v_i(t)v_j(0)\rangle$ by
\begin{equation}
\sigma_{ij}=\frac{m^*e^2}{\pi \hbar^2} \int_0^{\infty}\langle v_i(t)v_j(0)\rangle dt \label{eq:2}
\end{equation}
Here, the angle brackets denote averaging over all trajectories \cite{Richter2000}, and $m^*=0.067 m_e$ is the effective electron mass in GaAs. Since the formalism yields conductivities, its application to a single MB is not meaningful. Therefore, following Kubrak et al. \cite{Kubrak1999}, we first consider a periodic array of MBs, each given by eq. \eqref{eq:1}, and then calculate the resistance per barrier. This technique suffers from the difficulty of determining the appropriate value for the period $a$, which has been interpreted as an \emph{active length} \cite{Kubrak1999} of the MB. For the simulations, we used $a=2.5\,\mathrm{\mu m}$, i.e., such that one period contains 90\% of the integrated magnetic field of a single MB. The trajectories of $10^5$ electrons with the Fermi energy, injected in arbitrary directions at $y=0$ and arbitrary x within one lattice period, are calculated numerically to a length of $31\,\mathrm{\mu m}$. We have assumed small angle scattering with Gaussian distributed scattering angles (limited to the range $[-\pi,\pi]$) and Poisson distributed times between scattering events, with an expectation value equal to the measured $\tau_q$. The full width at half maximum of the scattering angle distribution of $0.157 \pi$ is given by the condition that the simulated value for $\tau_D$ has to agree with the experimentally determined one \cite{Cerchez2007}.

\begin{figure}
\includegraphics{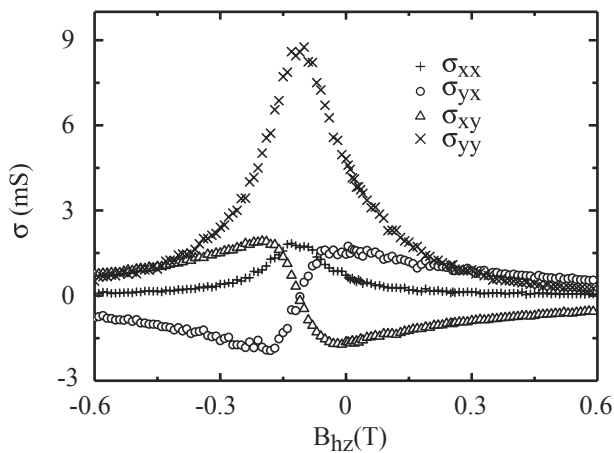}
\caption{Components of the conductivity tensor for a periodic array of MBs, as simulated within the Kubo model.} \label{fig:3}
\end{figure}

\begin{figure}
\includegraphics{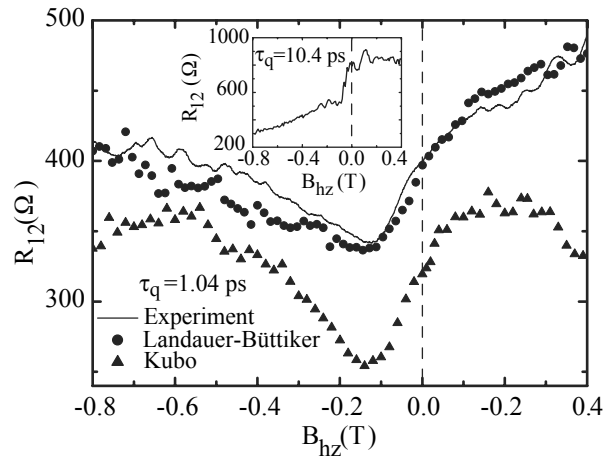}
\caption{Main figure: Comparison of the measured MB resistance as a function of $B_{hz}$ (full line) with the results of simulations based on the Kubo model (triangles) and on the Landauer-B\"uttiker model (circles). The inset shows a simulation of $R_{12}(B_{hz})$ for the quasi-ballistic case as obtained within the Landauer-B\"uttiker model. } \label{fig:4}
\end{figure}

In Fig. \ref{fig:3}, we show the simulated functions $\sigma_{ij}(B_{hz})$. They have shapes resembling those obtained within the Boltzmann model, from which they deviate in three respects. First, the $\sigma_{ij}(B_{hz})$ traces are displaced in $B_{z}$ - direction by $-\langle B_{fz}\rangle$, the average fringe field in one period, and thus centered at vanishing average magnetic field, $\langle B_{z}\rangle =0$. Second, the conductivities are scaled in $B_{hz}$ - and in $\sigma_{ij}$ - direction, with a particularly strong enhancement of $\sigma_{yy}$ as compared to $\sigma_{xx}$. This is a consequence of electrons moving in snake-type orbits along the MB in y-direction, as discussed in detail below. Finally, all $\sigma_{ij}$ are asymmetric around $\langle B_{z}\rangle =0$. This is most clearly visible in the off-diagonal elements, the extremal values of which for $\langle B_{z}\rangle >0$ are significantly smaller than those for $\langle B_{z}\rangle <0$. To compare these simulations with the measurements, we use
\begin{equation}
R_{12}=(\rho_{xx}-\rho_0)\frac{a}{W}+\frac{L}{W}\rho_0 \label{eq:3}
\end{equation}
where L denotes the distance between the voltage probes, W the width of the Hall bar and $\rho_0$ the resistivity of the 2DEG in zero magnetic field. $\rho_{xx}$ is given by
\begin{equation}
\rho_{xx}=\frac{\sigma_{yy}}{\sigma_{xx}\sigma_{yy}-\sigma_{xy}\sigma_{yx}} \label{eq:4}
\end{equation}
Fig. \ref{fig:4} shows that the simulated magnetoresistance, albeit too low, agrees reasonably with the measured one in terms of
magnitude and shape of the dip. The asymmetry is also visible, although less pronounced than in the experiments. The position of the simulated minimum corresponds to $\langle B_z\rangle =0$. We note that the
simulated trace of $\rho_{yy} (B_{hz})$ remains basically unaffected by the MBs, while $\rho_{xy}(\langle B_{z}\rangle )= -\langle B_{z}\rangle /ne$ (not shown). Moreover, the results for $\rho_{xx}$ are only weakly dependent on $a$ for $2\,\mathrm{\mu m}\leq a\leq 3\,\mathrm{\mu m}$.

It is well known that periodic, one-dimensional superstructures (the homogeneous direction being the y-direction) in 2DEGs can increase the diffusion in y-direction, which leads to a positive magnetoresistance $\rho_{xx}(B_z)$ around $B_z=0$ \cite{Beenakker1989a,Beton1990,Nogaret1997}. The situation in our experiments is different, since (i) $\langle B_{fz}\rangle  \neq 0$, and (ii) $B_{fz}(x)$ is asymmetric about $\langle B_z\rangle =0$. While the first point causes a simple displacement of $\rho_{xx}(B_{hz})$ in the $B_{hz}$ - direction, the second one generates asymmetries in $\rho_{xx}(B_{hz})$. In particular,
$-\sigma_{xy}\sigma_{yx} (|\langle B_z\rangle |)<-\sigma_{xy}\sigma_{yx} (-|\langle B_z\rangle | )$ which, according to eq. \eqref{eq:4}, increases $\rho_{xx}(|\langle B_z\rangle | )$ as compared to $\rho_{xx}(-|\langle B_z\rangle | )$. From this point of view, our experiments on single MBs are complementary to previous studies on periodic one-dimensional magnetic superlattices with vanishing average magnetic field, where both a positive magnetoresistance around B=0  and commensurability oscillations at higher magnetic fields have been found \cite{Ye1995,Nogaret1997}. Apparently, the positive magnetoresistance does not require a periodic structure and the related commensurate orbits, but rather emerges already from a single unit cell of the superlattice.

The quantitative differences from the measurements have several origins. In the simulated array, trajectories that are commensurate with the lattice do exist, but are absent in the experiments; the resistances of the MBs cannot be added according to Ohm's law as implied by eq. \eqref{eq:3}. Also, choosing a value for $a$ remains somewhat arbitrary. Furthermore, the simulation does not include the edges of the sample.

These discrepancies can be avoided with the multi-terminal Landauer-B\"uttiker formalism \cite{Buttiker1986,Beenakker1989}. We have modeled a sample with sizes as shown in Fig. \ref{fig:1}, including source and drain (which was set to zero potential) and voltage probes 1-4. The geometry with six contacts allows various consistency checks, e.g. simulations of the Hall effect outside the MB. The symmetry of the problem suggests the equivalence of the set of contacts $\mathrm{\{S,1,2\}}$ to $\mathrm{\{D,3,4\}}$.

We inject $1.8\times 10^5$ electrons at each cross-sectional line, defined at $2.5\,\mathrm{\mu m}$ inside the contacts S, 1 and 2. The electron positions on the injection lines and their injection angles are random, while their kinetic energy equals the Fermi energy. We solve the equations of motion for the electrons and calculate their trajectories until they pass the cross-sectional line of any of the six contacts. The sample edges are simulated as hard walls with specular reflectivity. Elastic scattering is included as described above, with scattering times adapted to the experiments as well as to a quasi-ballistic scenario.
Within the Landauer-B\"uttiker formalism, one finds
\begin{equation}
R_{12}=\frac{(V_1-V_2)}{I}=(M^{-1})_{31}-(M^{-1})_{21}
\label{eq:5}
\end{equation}
where
\[M= \left[ \begin{array}{ccccc}
G_S & -G_{S1} & -G_{S2} & -G_{S3} & -G_{S4}\\
-G_{1S} & G_1 & -G_{12} & -G_{13} & -G_{14}\\
-G_{2S} & -G_{21} & G_2 & -G_{23} & -G_{24}\\
-G_{3S} & -G_{31} & -G_{32} & G_3 & -G_{34}\\
-G_{4S} & -G_{41} & -G_{42} & -G_{4S} & G_4 \end{array} \right]\]

Here the notation

\begin{equation}
G_j=\sum_{i\neq j} G_{ji},\;j=S,1,2,3,4;\;i=S,1,2,3,4,D
\label{eq:6}
\end{equation}

has been used. The individual components of the conductance tensor are given by

\begin{equation}
G_{ji}=\frac{2e^2}{h}N_i\cdot T_{ji},\;j,i=S,1,2,3,4
 \label{eq:7}
\end{equation}

Here, $T_{ji}$ denotes the transmission probability from contact i to contact j, and $N_i$ is the number of modes in contact i.

The results of these simulations are reproduced in Fig. \ref{fig:4} in comparison to the experiment. We find good agreement after a constant resistance of $12\,\mathrm{\Omega}$ is added to the simulated trace of $R_{12}(B_{hz})$. This resistance originates from the parabolic magnetoresistivity of the 2DEG as a function of $B_{hx}$ \cite{Lee1985} used to magnetize the Dy film in the experiment \cite{Cerchez2007}, which at $B_{hx}\approx 8\,\mathrm{T}$ amounts to the value taken into account. We emphasize that there is no fit parameter in this simulation. The simulated curve reproduces all classical aspects of the measurement, namely the position of the minimum, the shape of the magnetoresistance dip, and the absolute values of the resistances. Similar agreement is obtained for other magnetizations, as indicated by the inset in Fig. 2(c), where the simulated $B_{hz}^{min}$ is compared with the measured ones. We tentatively attribute the remaining deviations to distortions in the sample not considered in the model, such as non-specular scattering and/or finite electric fields at the Hall bar edges, inhomogeneities of MB in y direction, the spatial variations of $B_{fx}$ as well as of the electron density across the MB, and thermal smearing.

\begin{figure}
\includegraphics{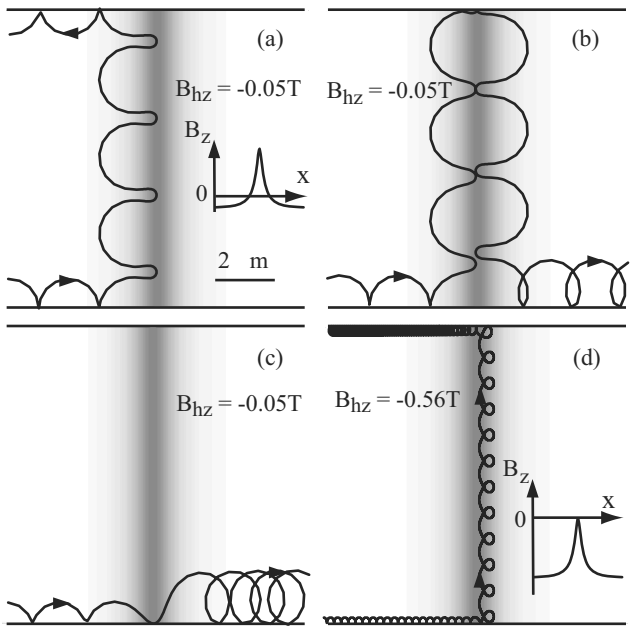}
\caption{Characteristic ballistic trajectories within the dip region, (a)-(c), as well as for $B_z(x)\leq 0$ for all x (d).  $B_{fz}(x)$ is indicated in gray scale, while the corresponding $B_z(x)$ is sketched in the insets.}
\label{fig:5}
\end{figure}

We proceed by specifying the electron trajectories responsible for the enhanced conductance. In Fig. \ref{fig:5}, four characteristic, ballistic trajectories starting from the source contact are shown for a MB with $B_{fz}^{max}=0.57\,\mathrm{T}$.

For $B_{hz}>-80\,\mathrm{mT}$, the MB is closed, i.e., transmission is only possible via $E\times B$ drift at the edges of the Hall bar. In this regime, three scenarios for electrons entering the MB region exist, shown in Figs. \ref{fig:5} (a)-(c). Depending on the initial condition, i.e. the injection angle of the electron into the Hall bar, the electron either moves in snake-type orbits along the MB in y-direction, see Figs. \ref{fig:5}(a,b), or crosses the MB at the lower edge of the Hall bar (we denote those orbits as edge orbits, Fig. \ref{fig:5}(c)). At the upper edge, the snake orbits split in two groups. They either get reflected or transmitted, depending on the point where they hit the upper edge which determines the direction of the $E\times B$ drift. As a result, a fraction of the trajectories that contribute to the enhancement of the conductance $S_{yy}$ also participates in the transmission through the MB, thereby coupling $S_{yy}$ to $S_{xx}$. Transport via edge orbits sets in at $B_{hz}=-50\,\mathrm{mT}$, and its weight increases as $B_{hz}$ is made more negative. The snake orbits, on the other hand, are present in the interval $-B_{fz}^{max} < B_{hz} <0$, in which $B_z (x)$ changes polarity. Their weights fluctuate, with an overall tendency to drop in the range where edge orbits exist. However, even for $B_{hz}\leq -B_{fz}^{max}$, the MB has an enhanced conductance in y direction, due to orbits of the type shown in Fig. \ref{fig:5}(d). Here, the magnetic field gradient at the MB leads to enhanced backscattering, which may be the reason why the Shubnikov-de Haas oscillations are suppressed at negative $B_{hz}$.

We have calculated $R_{12}(B_{hz})$ for a quasi-ballistic case with a rather high quantum scattering time, namely $\tau_q=10.4\,\mathrm{ps}$, which could nonetheless be realized experimentally with high mobility heterostructures. The result is shown in the inset of Fig. \ref{fig:4}. A \emph{step} rather than a dip in $R_{12}$ is obtained at the onset of edge orbits across the MB, at which the resistance drops by a factor of about 2. Moreover, in the regime where snake orbits are present, reproducible resistance fluctuations are found, which reflect the varying weight of those trajectories. A detailed discussion of these features is beyond the scope of this work. The simulations thus confirm that the increase of $R_{12}$ for $B_{hz}<-125\,\mathrm{mT}$ is caused by elastic scattering, an interpretation that is in tune with that one obtained within the Kubo model. We note that at lower mobilities, scattering between various snake orbits smears the simulated resistance fluctuations.\\

\section{\label{sec:4}SUMMARY AND CONCLUSIONS}
The resistance of a magnetic barrier as a function of superimposed, homogeneous perpendicular magnetic fields has been investigated both experimentally and theoretically. The magnetoresistance shows a characteristic asymmetric dip at a homogeneous magnetic field of opposite direction to that one of the barrier. Simulations show that electrons can travel along the barrier in snake orbits as well as in cyclotron-type orbits with magnetic-field gradient induced drifting centers. At the sample edges, edge orbits exist due to $E\times B$ drifts. They effectively couple $S_{yy}$ to $S_{xx}$, thereby reducing the resistance of the magnetic barrier. The observed asymmetric resistance dip is interpreted in terms of the combined effects of (i) enhancement of $S_{yy}$, (ii) transmission via edge orbits and (iii) elastic scattering in the barrier region. Simulations based on the multi-terminal Landauer-B\"uttiker model without adjustable parameters result in excellent agreement with the experiment. In addition, they predict parametric resistance fluctuations in the regime of small homogeneous perpendicular magnetic fields, which will be interesting to study experimentally in further work.

The authors acknowledge support of the Heinrich-Heine-Universit\"at D\"usseldorf. H. X. and T. H.gratefully acknowledge support by the Humboldt Foundation.


\end{document}